\documentclass[twocolumn,amsmath, prx,amssymb,superscriptaddress, longbibliography]{revtex4-2}

\usepackage{hyperref}
\hypersetup{
  colorlinks   = true, 
  urlcolor     = blue, 
  linkcolor    = blue, 
  citecolor   = blue 
}

\usepackage{xcolor}
\newcommand{\revise}[1]{\textcolor{blue}{#1}}

\usepackage{color}
\usepackage{graphicx}
\usepackage{dcolumn}
\usepackage{bm}
\usepackage{diagbox}

\DeclareUnicodeCharacter{2212}{-}

\begin{document}
\preprint{APS/123-QED}

\title{Hybrid Gate-Based and Annealing Quantum Computing for Large-Size Ising Problems}

\author{Chen-Yu Liu}
\email{d10245003@g.ntu.edu.tw}
\affiliation{Graduate Institute of Applied Physics, National Taiwan University, Taipei 10617, Taiwan}
\author{Hsi-Sheng Goan}
\email{goan@phys.ntu.edu.tw}
\affiliation{Graduate Institute of Applied Physics, National Taiwan University, Taipei 10617, Taiwan}
\affiliation{Department of Physics and Center for Theoretical Physics, National Taiwan University, Taipei 10617, Taiwan}
\affiliation{Center for Quantum Science and Engineering, National Taiwan University, Taipei 10617, Taiwan}
\affiliation{Physics Division, National Center for Theoretical Sciences, Taipei, 10617, Taiwan}

\begin{abstract}

One of the major problems of most quantum computing applications is that the required number of qubits to solve a practical problem is much larger than that of today's quantum hardware. Therefore, finding a way to make the best of today's quantum hardware has become a critical issue.
In this work, we propose an algorithm, called large-system sampling approximation (LSSA), to solve Ising problems with sizes up to $N_{\text{gb}}2^{N_{\text{gb}}}$ by an $N_{\text{gb}}$-qubit gate-based quantum computer, and with sizes up to $N_{\text{an}}2^{N_{\text{gb}}}$ by a hybrid computational architecture of an $N_{\text{an}}$-qubit quantum annealer and an $N_{\text{gb}}$-qubit gate-based quantum computer. 
By dividing the full-system problem into smaller subsystem problems, the LSSA algorithm then solves the subsystem problems by either gate-based quantum computers or quantum annealers, and optimizes the amplitude contributions of the solutions of the different subsystems with the full-problem Hamiltonian by the variational quantum eigensolver (VQE)  on a gate-based quantum computer. After optimizing the VQE amplitude contributions, the approximated ground-state configuration, which is the approximated solution of a corresponding quadratic unconstrained binary optimization (QUBO) problem, will be determined. 
We apply the level-1 approximation of LSSA to solving fully-connected random Ising problems up to 160 variables using a 5-qubit gate-based quantum computer, and solving portfolio optimization problems up to 4096 variables using a 100-qubit quantum annealer and a 7-qubit gate-based quantum computer. Moreover, LSSA can be further extended to 
a deeper level of approximation. We demonstrate the use of the level-2 approximation of LSSA to solve the portfolio optimization problems up to 5120 ($N_{\text{gb}}2^{2N_{\text{gb}}}$) variables with pretty good performance by using just a 5-qubit ($N_{\text{gb}}$-qubit) gate-based quantum computer. The effects of different subsystem sizes, numbers of subsystems, and full problem sizes on the performance of LSSA are investigated on both simulators and real hardware. 
The completely new computational concept of the hybrid gate-based and annealing quantum computing architecture opens a promising possibility to investigate large-size Ising problems and combinatorial optimization problems, making practical applications by quantum computing possible in the near future.

\end{abstract}

\maketitle 


\section{Introduction}

Accessible cloud quantum computing resources provided by several companies \cite{ibmquantum, awsbraket, dwave} makes the development of quantum computing applications practical and popular in recent years. Appropriate mapping from the original problem to quantum solvable form is required to solve problems on quantum computers. A popular choice for solving a large class of combinatorial optimization problems that can be converted into quadratic unconstrained binary optimization (QUBO) problems is mapping the problem to an Ising Hamiltonian and solving for the corresponding ground state of the Hamiltonian, which is related to the solution of the original problem \cite{toising1, toising2}. 
Applications of quantum optimization to real-world problems have been demonstrated for portfolio optimizations \cite{portfolio2, portfolio3, portfolio4, portfolio5}, industrial optimization problems \cite{supplychain1, industry1}, and traveling salesman problems \cite{tsp1, tsp2}. To find the optimal solution of the corresponding Hamiltonian, the 
gate-based quantum computing with variational quantum eigensolver (VQE) uses parameterized gates to construct a trial state and optimizes the best set of parameters that can approximate the ground state of the Ising Hamiltonian \cite{vqe1}. The quantum approximate optimization algorithm (QAOA) is one of the branches of VQE, where the trotterization of the adiabatic evolution makes it a scalable approach \cite{qaoa1}. However, VQE and QAOA are designed to solve the Ising Hamiltonian problems with $N_p$ variables on an $N_p$-qubit gate-based quantum computer. Thus even the largest gate-based quantum computer to date provided by IBM (\textsf{ibm\_washington}) can only solve the problem with 127 variables if we use the original VQE and QAOA algorithms. 

To face this issue, several qubit-efficient methods are proposed to tackle the problem with size larger than the hardware size. Divide-and-Conquer QAOA (DC-QAOA) 
that was proposed for graph maximum cut (MaxCut) problems
divides a large graph recursively to the size of the available hardware 
with each step splitting a resultant graph into exactly two subgraphs \cite{dcqaoa1}. Then a classical method of  the combination policy of quantum state reconstruction is used to reconstruct the solution by inputting the sampling distribution of each subgraph solution. However, DC-QAOA only applies to input graphs with connectivity less than maximum allowed qubit size, which makes it probably inapplicable to the fully-connected graph problems \cite{dcqaoa1}. 
Additionally, QAOA-in-QAOA, also divides and conquers the graph MaxCut problem by utilizing the structure of graphs and the $\mathbb{Z}_2$ symmetry, and after the subgraphs are solved, 
the merging of local solutions of all subgraphs is reformulated into a new Maxcut problem \cite{qaoainqaoa1}. Although maximally 9-degree graphs are tested in the paper, the applicability and the performance on problems other than MaxCut, e.g., on the complete (fully-connected) graphs, are not discussed \cite{qaoainqaoa1}.
Another method of cluster-VQE shows an example in quantum chemistry that splitting qubits into different clusters and minimizing the correlations between clusters, with the help of a method called ``Hamiltonian dressing''. But depends on the problem Hamiltonian (strongly correlated systems, for example), computational scaling of the algorithm can be exponential with respect to the dressing steps \cite{clustervqe1}.
Besides, an encoding scheme is also proposed to find the solution by smaller hardware \cite{qbeff1}. 


These approaches focus on using gate-based quantum computers, while a quantum annealer which usually has a larger qubit number is also a candidate to solve the large-size Ising problem. As an example of that, D-Wave also provides a quantum-classical hybrid solver to break the full problem into small pieces of subsystem problems and then recombine the solutions of the solved subsystem problems classically to obtain an approximate solution for the full problem \cite{qbsolv1, qbsolv2}. In most cases, the sizes of subsystem problems solved by gate-based chips are much smaller than those by annealing chips, while gate-based chips can utilize gate operations to encode information that may be helpful in a problem-solving scheme.

Combining the advantages of gate-based and annealing quantum computing approaches mentioned above, we propose a hybrid structure that can solve the subsystem problems by either gate-based chips or annealing chips, and then the solutions of the subsystem problems are recombined with amplitudes that can be optimized by VQE on gate-based chips. Our proposed algorithm can solve fully-connected random Ising problems that are $O(10^0)$ and portfolio optimization problems that are $O(10^1)$ larger in size than the available quantum annealers and gate-based quantum computers, with pretty good performance of the results from both simulator and real hardware. With the polynomial time scaling behavior and capability of solving subsystems in parallel, this algorithm could be feasible for large-size real world problem using today's quantum hardware. 

This paper is organized as follows.
we will first introduce the idea of the proposed algorithm in Sec.~\ref{sec:lssa} and Sec.~\ref{sec:ao}. In Sec.~\ref{sec:eds}, we then use the method of exact diagonalization on classical computers and quantum simulators to systematically investigate how the approximation ratio changes for different parameters of the proposed algorithm. After investigating the problem on quantum simulators, we will show further result obtained from real hardware in Sec.~\ref{sec:gbcqa}. We will discuss the results in Sec.~\ref{sec:discuss}, along with the proper use cases for different kinds of problems. Then we will summarize the paper and future work based on the results in Sec.~\ref{sec:outlook}.



\begin{figure*}[htp]
\centering
\includegraphics[scale = 0.35]{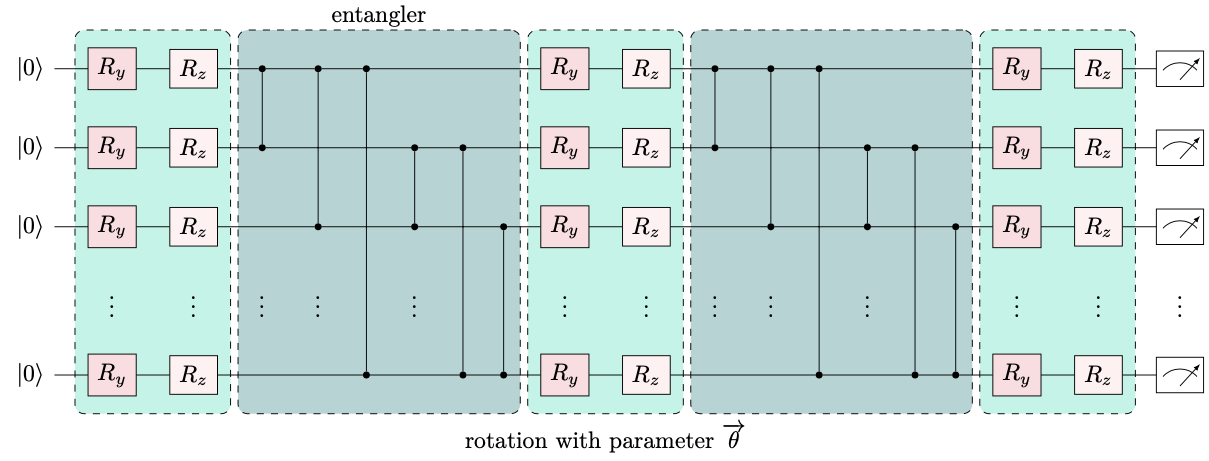}
\caption{Quantum circuit structure of VQE for subsystem amplitude optimization. 
The initial state $|0 \rangle^{\otimes N_{\text{gb}}}$ is followed by a circuit of alternating rotation layer and entanglement layer. The rotation layer is composed of parameterized single-qubit rotation gates and the entanglement layer is composed of two-qubit entangling gates.
Here, in the case of a $N_{\text{gb}}$-qubit quantum circuit with two entanglement layers sandwiched by three rotation layers, a total of $6N_{\text{gb}}$ tunable parameters of the rotation angles of the $R_y$ and $R_z$ rotation gates is denoted as $\vec{\theta} = (\theta_1, \theta_2, \cdots, \theta_{6N_{\text{gb}}})$.} 
\label{fig:vqe}
\end{figure*}

\section{Large-System Sampling Approximation}
\label{sec:lssa}

Several classes of QUBO problems can be mapped into Ising problems by the concept that the binary variable $x_i \in \{0,1\}$ can be related to the up and down spin state  in a spin$-\frac{1}{2}$ system by $z_i = 2 x_i - 1$, where $z_i \in \{-1,1\}$ is the spin variable. Variable $x_i$ could represent the binary decision, the label of the node in a graph, or even the binary bit in a factorization problem \cite{toising1, toising2}. Making use of the superposition nature of the quantum system, the spin system can then be solved by quantum computers. In this work, we consider Ising problems in the form: 
\begin{equation}
\label{Isingproblem}
H = \sum_{i, j =1 }^{N_p} J_{ij} z_i z_j + \sum_{i=1}^{N_p} h_i z_i,
\end{equation}
where $z_i$ are the spin variables, $N_p$ is the number of spin variables (or the problem size), and $h_i$ and $J_{ij}$ correspond respectively to the bias and coupling strength (weight) of the spin system with indices 
$i,j \in \{1,2,\cdots, N_p\}$. Here the values of $h_i$ and $J_{ij}$ are considered to be in the range of $(-1, 1)$.

Instead of solving the problem of the full $N_p$-spin system directly, we try to group the spins into smaller subsystems and then solve the problems of the subsystems first. After collecting the results of the subsystems, an estimation of the full system result will be made based on the statistical behavior of the subsystem results. Since this method is intended to approximate the solution of a large-scale full-system problem by sampling various subsystems with solvable problem size and then optimizing the amplitude contributions of the  the subsystem solutions through the full-system Hamiltonian, we called it the large-system sampling approximation (LSSA). For an $N_p$-spin Ising problem with the Hamiltonian of Eq.~(\ref{Isingproblem}), consider a sampling method (described later) that picks a subsystem of $N_g$ sites 
with $N_g \le N_p$, and then the subsystem Hamiltonian can be written as:
\begin{equation}
H_{\text{sub}} = \sum_{i',j'=1}^{N_g} J_{i' j'} z_{i'} z_{j'} + \sum_{i'=1}^{N_g} h_{i'} z_{i'},
\end{equation}
where the subsystem sites have been labeled with indices $i'$ and $j'$, where $i', j' \in \{1,2,\cdots, N_g\}$.
After we solved the relatively simple (compare to $H$) eigenvalue problem of $H_{\text{sub}}$, the corresponding ground state $|\text{GS}_{\text{sub}} \rangle = [s_{1}, s_{2},\cdots, s_{{N_g}}]^{T}$ is obtained, where spin variables $s_{i'} \in \{-1,1 \}$. Next, we relabel the subsystem labels $i', j'$ back to the full-system labels $i,j$, and thus obtain a part of the spin configuration of the full system, that is, a vector of length $N_p$ with $N_g$ non-empty elements.
For the subsystem ground state obtained from the $i$th sampled subsystem Hamiltonian, we label it as $|\text{GS}_{\text{sub}} ^{(i)}\rangle$. The subsystem sampling will be performed $N_s$ times in the following manner. The subsystems of size-$N_g$ will be picked randomly and non-repetitively until all of the variables (sites) in a full problem are picked at least once. After that, if the number of the subsystems is below $N_s$, the next round of non-repetitive random sampling will be made until all of the variables are picked at least twice or until the number of subsystems is $N_s$. The sampling procedure will be carried out in this fashion until all $N_s$ subsystems are formed, and thus all the variables will be picked at least $\lfloor (N_s \times N_g)/N_p \rfloor$ times. The condition $N_s \times N_g \ge N_p$ must be satisfied to cover all the variables in a problem at least once.

After the results of $N_s$ subsystems are obtained, a vector $| \mathcal{S^{\text{wc}}} \rangle$ of the weighted subsystem ground states
\begin{equation}
\label{swc}
|\mathcal{S^{\text{wc}}}\rangle = \sum_{i =1 }^{N_s} C^{(i)} |\text{GS}_{\text{sub}} ^{(i)}\rangle, 
\end{equation}
is constructed by summing the result of the ground state of each sampling $|\text{GS}_{\text{sub}} ^{(i)}\rangle$ weighted by some undetermined coefficient $C^{(i)}$, denoting the contribution of each subsystem, which will be discussed and determined in the next section. In this stage, we have a statistical view of the spin configuration after sampling subsystems $N_s$ times, and furthermore, an approximation of the ground state configuration of the full-system $H$ is determined by the sign of each element in $|\mathcal{S^{\text{wc}}} \rangle$: 
\begin{equation}
\label{signswc}
|\text{sign}(\mathcal{S^{\text{wc}}})\rangle \approx |\text{GS}_{\text{full}} \rangle,
\end{equation}
which represent a new vector where the $i$th element  $|\text{sign}(\mathcal{S^{\text{wc}}})\rangle_i$ is $+1$ for $|\mathcal{S^\text{wc}}\rangle_i \ge 0$ and is $-1$ for $|\mathcal{S^\text{wc}}\rangle_i < 0$. It is expected that $|\text{sign}(\mathcal{S^{\text{wc}}})\rangle$ will reduce to $|\text{GS}_{\text{full}} \rangle$ when the subsystem size approaches the full-system size, i.e., $N_g \rightarrow N_p$. 

As a simple example to illustrate how to obtain $|\text{sign}(\mathcal{S^{\text{wc}}})\rangle$ in LSSA, let us consider a problem of  
size $N_p = 4$, subsystem size $N_g = 2$, and the number of subsystems $N_s = 4$.
Suppose the results of the solved subsystem ground states are 
\begin{eqnarray}
&&| \text{GS}_{\text{sub}}^{(1)} \rangle = [1, -1, 0, 0], \label{GS1}\\
&&| \text{GS}_{\text{sub}}^{(2)} \rangle = [0, 0, 1, -1],  \\
&&| \text{GS}_{\text{sub}}^{(3)} \rangle = [-1, 0, 0, 1],  \\
&&| \text{GS}_{\text{sub}}^{(4)} \rangle = [0, 1, 1, 0]. \label{GS4}
\end{eqnarray}
Note that the elements with value $0$ in Eqs.~(\ref{GS1})-(\ref{GS4}) corresponds to the empty information of the site that is not selected in the sampled subsystem configuration. 
In this case, $|\mathcal{S^{\text{wc}}}\rangle $ from Eq.~(\ref{swc}) becomes
\begin{eqnarray}
    |\mathcal{S^{\text{wc}}}\rangle &=& [C^{(1)}-C^{(3)}, -C^{(1)}+C^{(4)}, \nonumber \\
    &&C^{(2)}+C^{(4)}, -C^{(2)}+C^{(3)}]. 
\end{eqnarray}
After the VQE optimization procedure to determine the values of the amplitudes $C^{(i)}$ described in the next section, suppose we have $C^{(1)}-C^{(3)} > 0$, $-C^{(1)}+C^{(4)} < 0$, $C^{(2)}+C^{(4)} > 0$, and $-C^{(2)}+C^{(3)} < 0$. Then from Eq.~(\ref{signswc}), the approximated ground state reads $|\text{GS}_{\text{full}}\rangle \approx  |\text{sign}(\mathcal{S^{\text{wc}}})\rangle = [1, -1, 1, -1]$.
 




\begin{figure*}[htp]
\centering
\includegraphics[scale=0.29]{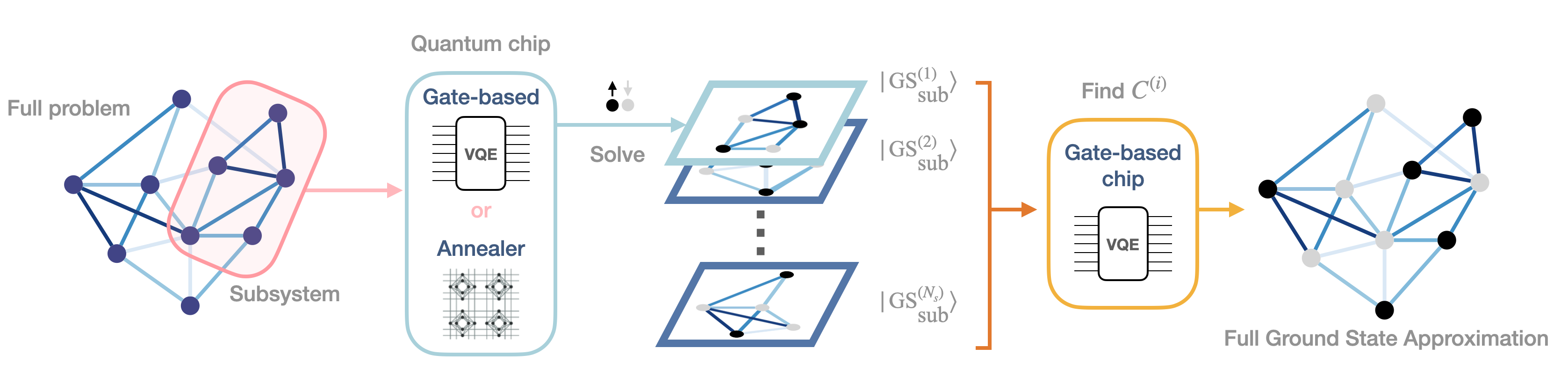}
\caption{Schematic flow chart of LSSA. After a procedure of random grouping of the spin variables from the full-system problem to form subsystems, the subsystem ground-state problems are solved by gate-based or annealing quantum chips. The resulting subsystem ground states are used to construct a trial state vector $|\mathcal{S}^{\text{wc}}\rangle$ with undetermined coefficients (amplitudes) described in Eq.~(\ref{swc}), and then VQE is used to  find the optimal coefficients with the cost function of Eq.~({\ref{cost}}).} 
\label{fig:flow}
\end{figure*}

\section{Amplitude Optimization}
\label{sec:ao}
In the previous section, we constructed a vector $|\mathcal{S}^{\text{wc}}\rangle$ by Eq.~(\ref{swc}), which has a set of undetermined coefficients or amplitudes $C^{(i)}$. These amplitudes can be considered as a set of tunable parameters, and one can determine the contribution of each amplitude $C^{(i)}$ from $| \text{GS}_{\text{sub}}^{(i)} \rangle$ of the $i$th subsystem configuration by minimizing the full-system ground state energy
using the full-system Hamiltonian $H$ with the VQE algorithm. To encode $N_s$ coefficients into the amplitudes of the quantum state, $N_{\text{gb}} = \lceil \log_2 N_s \rceil$ qubits of a gate-based quantum system is required, 
and the cost function of the VQE algorithm is defined as: 
\begin{equation}
\label{cost}
\text{Cost}(\mathbf{C}(\vec{\theta})) = \langle \text{sign}(\mathcal{S}^{\text{wc}}) | H | \text{sign}(\mathcal{S}^{\text{wc}}) \rangle,
\end{equation}
where $\mathbf{C}(\vec{\theta}) = (C^{(1)}(\vec{\theta}), C^{(2)}(\vec{\theta}), \cdots, C^{(N_s)}(\vec{\theta}))$ 
and $\vec{\theta}$ represents a set of tunable parameters in the parameterized quantum circuit in the VQE algorithm. 
The structure of the quantum circuit of our VQE consists of alternating rotation layer and entanglement layer. The rotation layer is composed of single-qubit rotation gates and the entanglement layer is composed of two-qubit entangling gates.
A typical quantum circuit shown in Fig.~\ref{fig:vqe} contains two entanglement layers sandwiched by three rotation layers.
The parameterized quantum circuits we use in the VQE are the rotation layer circuits constructed by $R_y$ and $R_z$ rotation gates with a total of $6N_{\text{gb}}$ tunable parameters of the rotation angles, that is, $\vec{\theta} = (\theta_1, \theta_2, \cdots, \theta_{6N_{\text{gb}}})$.   Additionally, it is also possible to have a different circuit structure of VQE, with different numbers of layers, configurations of gate connectivity and types of rotation and entangling gates, which can all be considered as hyperparameters that can be varied in VQE.
Appendix \ref{sec:eodvqe} shows the effect of these hyperparameters on our results. In this formulation, a single $N_{\text{gb}}$-qubit gate-based quantum computer can solve QUBO problems up to $N_{\text{gb}} 2^{N_{\text{gb}}}$ variables as each subsystem size can have up to $N_{\text{gb}}$ variables to be solved by algorithms like QAOA, and maximally $2^{N_{\text{gb}}}$ subsystems can be encoded into the VQE for amplitude optimization. Interestingly, 
it is possible to use a quantum annealer to solve a significantly larger size of subsystem problems and then encode them to a gate-based quantum computer for subsystem amplitude optimization.
In this case, a QUBO problem with size up to $N_{\text{an}}2^{N_{\text{gb}}}$ variables could be handled, where $N_{\text{an}}$ is the size of the subsystems that can be solved by the quantum annealer. The corresponding results will be discussed later. The flow of LSSA including amplitude optimization is shown in Fig.~\ref{fig:flow}. In this paper, $\vec{\theta}$ are updated by the COBYLA optimization method up to 200 iterations.


\section{Exact Diagonalization and Simulators} 
\label{sec:eds}

\subsection{Random Ising Problems}

\begin{figure}[htp]
\centering
\includegraphics[scale=0.25]{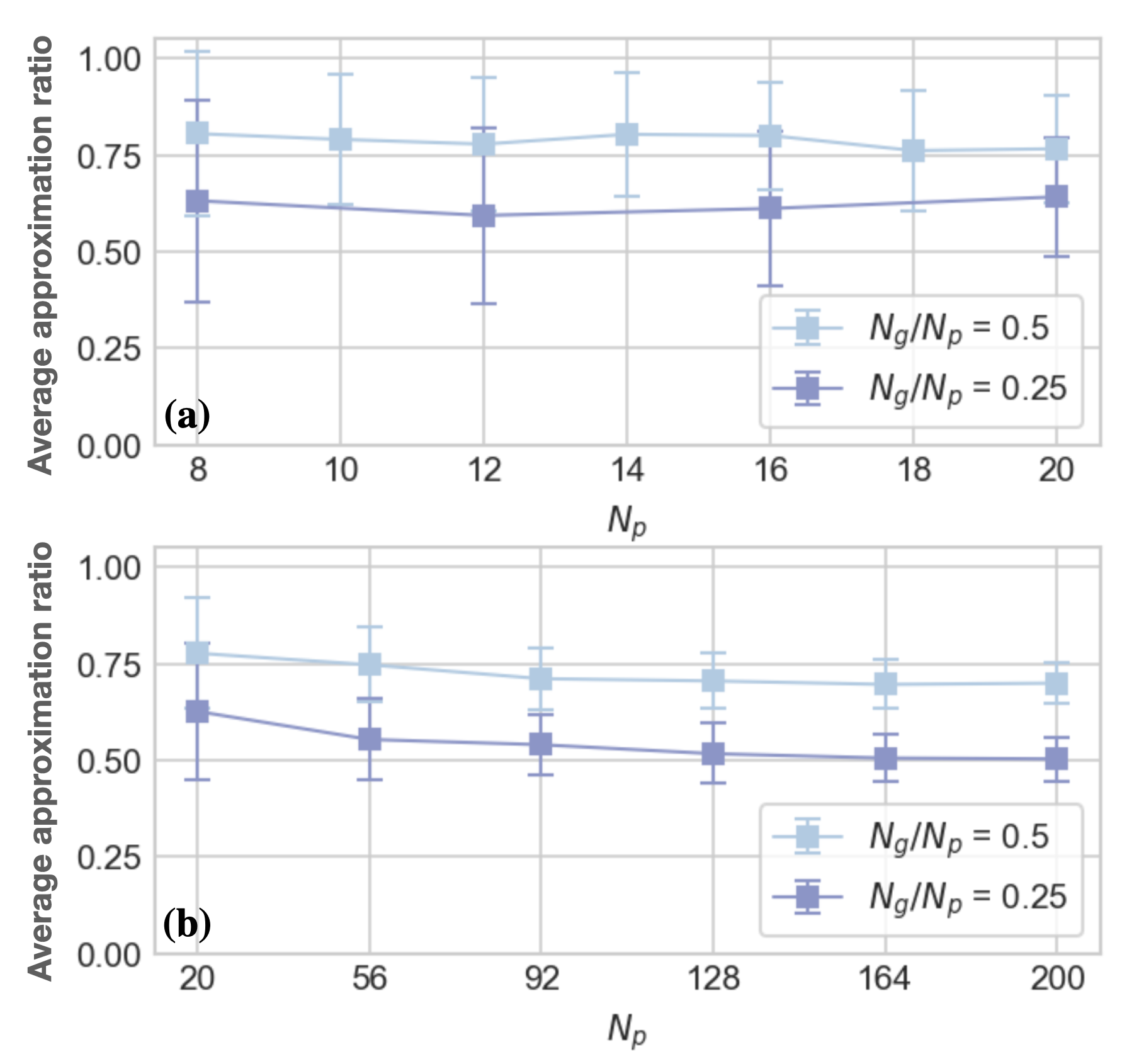}
\caption{(a) Average approximation ratio for the random 
Ising problems with different values of $N_p$
from 8 to 20. 
For each $N_p$, 100 random Ising Hamiltonians are tested and each is solved by sampling respectively size-$N/2$ subsystems and size-$N/4$ subsystems $\frac{2N_p}{N_g}$ times. The subsystem problems are solved by exact diagonalization and the amplitude optimization is done by \textsf{ibmq\_qasm-simulator}. The approximation ratio is defined as  $\frac{\textrm{LSSA GSE}}{\textrm{Exact GSE}}$ here. (b) Average approximation ratio for the random Ising problems with different values of $N_p$ from 20 to 200. Other parameters are Similar to (a) except that the subsystem problems are solved by the \textsf{Dwave-tabu} classical solver and the approximation ratio here is defined as $\frac{\textrm{LSSA GSE}}{\textsf{Dwave-tabu }  \textrm{GSE}}$.}
\label{fig:simu_ar_1}
\end{figure}

In this paper, all the random Ising problems considered are fully connected unless we specifically mention. First, we test the LSSA algorithm for the random Ising problems in the form of Eq.~(\ref{Isingproblem}) with problem sizes from $N_p = 8$ to $N_p = 20$. For each problem size, 100 random Ising Hamiltonians are constructed and each is solved by sampling subsystems $\frac{2N_p}{N_g}$ times with subsystem size $N_g=N_p/2$ and $N_g=N_p/4$, respectively.
The ranges of the random couplings $J_{i j}$ and the biases $h_i$ are both constrained into the range $(-1,1)$. 
We use the approximation ratio $R_\textrm{ar}$, defined as the ration of the ground state energy (GSE) obtained by LSSA to GSE obtained by the best affordable method, 
as an indicator for performance. The higher of the value of $R_\textrm{ar}$ approaches 1, the better performance of LSSA is for the problem considered.
For the random Ising problems with sizes from $N_p = 8$ to $N_p = 20$,
the ground state and ground state energy of the full system and subsystems can be calculated by exact diagonalization in reasonable time, so both the full system and subsystem problems are solved by exact diagonalization while the amplitude optimization part for the subsystems' contributions is done on \textsf{ibmq\_qasm-simulator}.
In this case the approximation ratio is  $ R_\textrm{ar} \equiv \frac{\textrm{LSSA GSE}}{\textrm{Exact GSE}}$. 
The corresponding result is shown in Fig.~\ref{fig:simu_ar_1}(a).

To explore the behavior of LSSA for problems with larger system sizes, from $N_p =20$ to $N_p =200$, we use the \textsf{Dwave-tabu} solver as our classical simulator,  
and the approximation ratio in this case is defined as $R_\textrm{ar} \equiv \frac{\textrm{LSSA GSE}}{\textsf{Dwave-tabu } \textrm{GSE}}$.
The corresponding result in Fig.~\ref{fig:simu_ar_1}(b)  
shows how LSSA performs when sampling subsystems with sizes $N_g=N_p/2$ and $N_g=N_p/4$, randomly $N_s=\frac{2N_p}{N_g}$ times to solve the full problem with size $N_p$, compared to directly solving the size-$N_p$ system by \textsf{Dwave-tabu}.
As the sampling number of the subsystems is fixed to $N_s=\frac{2N_p}{N_g}$ for each $N_p$, 
so $N_s=4$ for $N_g=N_p/2$ and $N_s=8$ for $N_g=N_p/4$.
Thus a smaller fraction of possible combinations of subsystems is sampled when the system size $N_p$ is increased, i.e., only a fraction of $N_s/(\frac{N_p!}{N_g!(N_p-N_g)!})$ is sampled.
So a decreasing trend in the average approximation ratio can be seen in Fig.~\ref{fig:simu_ar_1}(b) when increasing the system size $N_p$ from 20 to 200. 
The average approximation ratio, however, fluctuates a little bit with $N_p$ as shown in Fig.~\ref{fig:simu_ar_1}(a), indicating the dependence on the spin variable configurations of the 4 (for $N_g=N_p/2$) or 8 (for $N_g=N_p/4$) randomly selected subsystems for smaller values of $N_p$ from 8 to 20.  
By considering the Hilbert space dimension increases exponentially with the system (subsystem) size, even though we only sample quite a small fraction of the possible combinations of subsystems (e.g., $4/(\frac{200!}{100!100!})$ for $N_p = 200$, $N_g=N_{p}/2$ and $N_s = 4$  subsystems), the average of the approximation ratio with 100 random weighted Ising Hamiltonians can still reach $\approx 68\%$.
One can also see from Figs.~\ref{fig:simu_ar_1}(a) and \ref{fig:simu_ar_1}(b) that results  for $N_g=N_p/2$ have a better average approximation ratio than those for $N_g = N_p/4$ as expected.


\begin{figure*}[htp]
\centering
\includegraphics[scale=0.33]{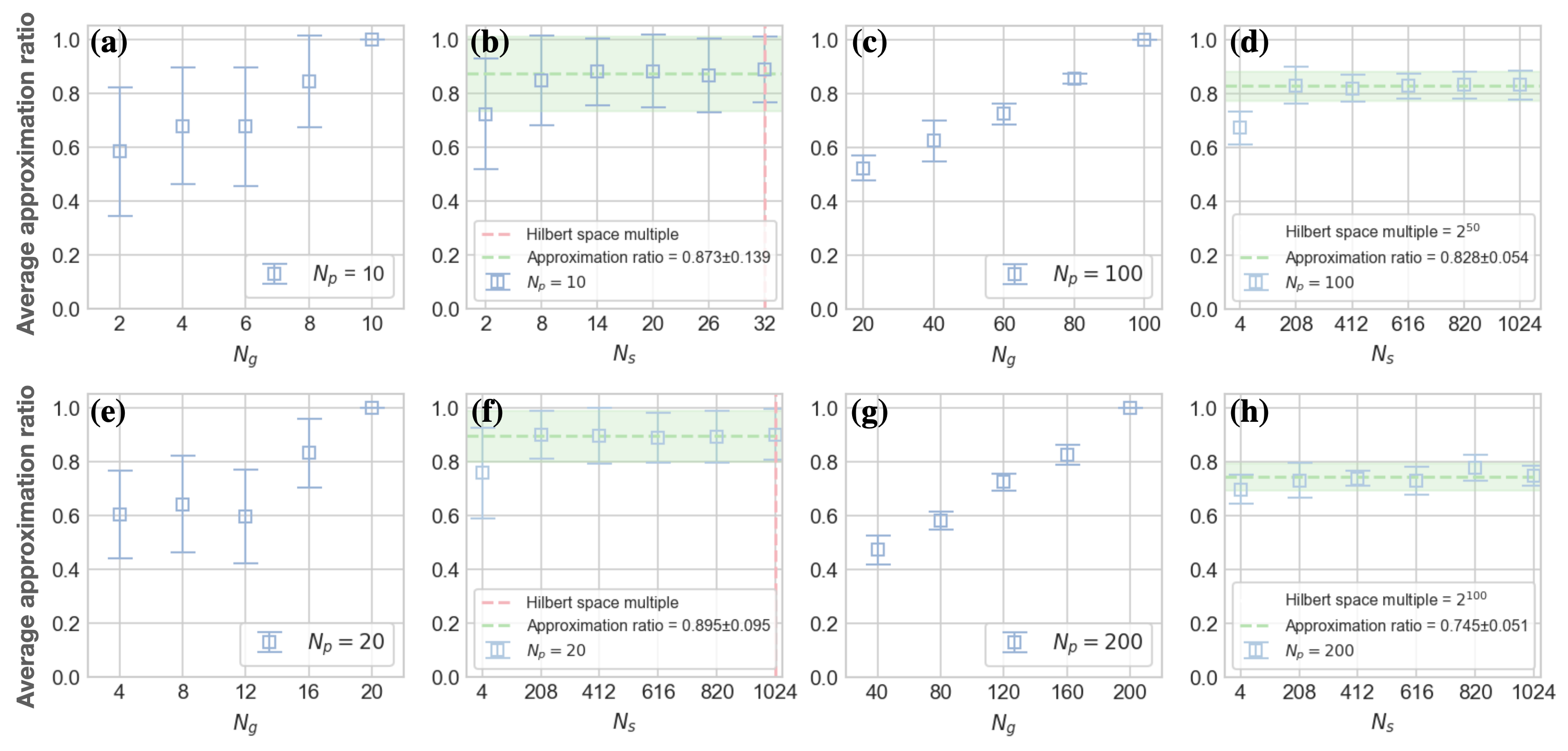}
\caption{Average approximation ratio for the random Ising problems with different subsystem sizes $N_g$ in (a), (e), (c) and (g), and different numbers of the sampled subsystems $N_s$ in (b), (f), (d) and (h) for problem sizes $N_p = 10, 20, 100$ and $200$, respectively. In (a), (c), (e), and (g), $N_s$ is fixed to 4, and in (b), (d), (f), and (h), $N_g$ is fixed to $N_p/2$. For each data point in (a), (b), (e), and (f) 100 random Ising problems are used 
and in (c), (d), (g), and (h), 10 random Ising problems are used.
The Hilbert space multiples $2^{N_p / 2}$ are shown by the pink dashed line in (b) and (f), and are mentioned in values in the legends of (d) and (h).}
\label{fig:simu_ar_2}
\end{figure*}

The effects of different subsystem sizes $N_g$ and numbers of sampled subsystems $N_s$ on the average approximation ratio are investigated in Fig.~\ref{fig:simu_ar_2}. The subsystem sizes $N_g$ of random Ising problems are tuned from a small size to the size of the full problem $N_p$ in Fig.~\ref{fig:simu_ar_2}(a) for $N_p = 10$ and in Fig.~\ref{fig:simu_ar_2}(c) for $N_p = 20$  with $N_s$ fixed to 4. 
As expected, the average approximation ratio $R_\textrm{ar} \rightarrow 1$ when $N_g \rightarrow N_p$ since more couplings between spins are considered when increasing $N_g$ until the full problem is considered.
For $N_p = 10$ in  Fig.~\ref{fig:simu_ar_2}(b) and $N_p = 20$ in Fig.~\ref{fig:simu_ar_2}(d),
 we tuned $N_s$ from a small number to the Hilbert space multiple with the size of a subsystem $N_g$ fixed to the value of $N_p/2$. The Hilbert space multiple is defined as the multiple between the dimension of the full problem Hilbert space and that of the subsystem, so the Hilbert space multiple here is $2^{N_p} / 2^{N_g} = 2^{N_p /2 }$. We observed that the average approximation ratio approaches a steady value 
and no longer increases when $N_s$ is larger than some value that is quite smaller than the Hilbert space multiple. 
To be more specific, the average approximation ratio $\approx 0.873 $ as $N_s \gtrapprox 11$ for the $N_p = 10$ case and the average approximation ratio $\approx 0.895$ as $N_s \gtrapprox 208$ for the $N_p = 20$ case.
To investigate the effects of different values of $N_g$ and $N_s$ in larger system sizes $N_p$, we consider the cases for $N_p = 100$ and $N_p=200$ with 10 random Ising problems tested for each data point of the average approximation ratio with fixed values of $N_g$ and $N_s$. 
Note that the definitions of the approximation ratio 
in Figs.~\ref{fig:simu_ar_2}(a), \ref{fig:simu_ar_2}(b), \ref{fig:simu_ar_2}(e), and \ref{fig:simu_ar_2}(f) is the same as that in Fig.~\ref{fig:simu_ar_1}(a), while 
in Figs.~\ref{fig:simu_ar_2}(c), \ref{fig:simu_ar_2}(d), \ref{fig:simu_ar_2}(g), and \ref{fig:simu_ar_2}(h), it is the same as that in Fig.~\ref{fig:simu_ar_1}(b).
That is,  the denominator of the approximation ratio in the former case is ``Exact GSE'' while it is ``\textsf{Dwave-tabu} GSE'' for the latter case.  
The behaviors of the average approximation ratio as the subsystem size $N_g$ is tuned up from a smaller value to the value of $N_p$ with a fixed subsystem size number $N_s = 4$  for $N_p = 100$ and $N_p = 200$ are 
shown in Figs.~\ref{fig:simu_ar_2}(c) and \ref{fig:simu_ar_2}(g), respectively. 
As expected, the behavior that  the average approximation ratio $\rightarrow 1$ as $N_g \rightarrow N_p$ is also observed. 
Similar to Fig.~\ref{fig:simu_ar_2}(f), $N_g$ is fixed to $N_p /2$ for $N_p = 100$ in Fig.~\ref{fig:simu_ar_2}(d) and $N_p = 200$ in Fig.~\ref{fig:simu_ar_2}(h), while $N_s$ is tuned up from $4$ to $1024$.
Note that the data points are only up to $N_s=1024$ as the data for $N_s$ approaching the Hilbert space multiple in these cases is too large to be sampled ($2^{50}$ and $2^{100}$). Nevertheless, in the case of $N_p = 100$ in Fig.~\ref{fig:simu_ar_2}(d), the average approximation ratio still converge to around $0.828$ as $N_s \gtrapprox 208$. In the $N_p = 200$ case, the average approximation ratio grows very slowly with the average value of last $5$ data points being $0.745$, so we may assume that the average approximation ratio will converge to some value with a sufficient large $N_s$ as in previous cases.
The results in Figs.~\ref{fig:simu_ar_2}(b), \ref{fig:simu_ar_2}(d), \ref{fig:simu_ar_2}(f), \ref{fig:simu_ar_2}(h) show that the minimal $N_s$ value requirement for LSSA to obtain a considerably good convergent performance for a fixed $N_g$ value is quite small compared to the size of the Hilbert space of the full problem. To further increase the performance of the LSSA algorithm, one may increase the value of the subsystem size $N_g$.

\begin{figure*}
\includegraphics[scale=0.33]{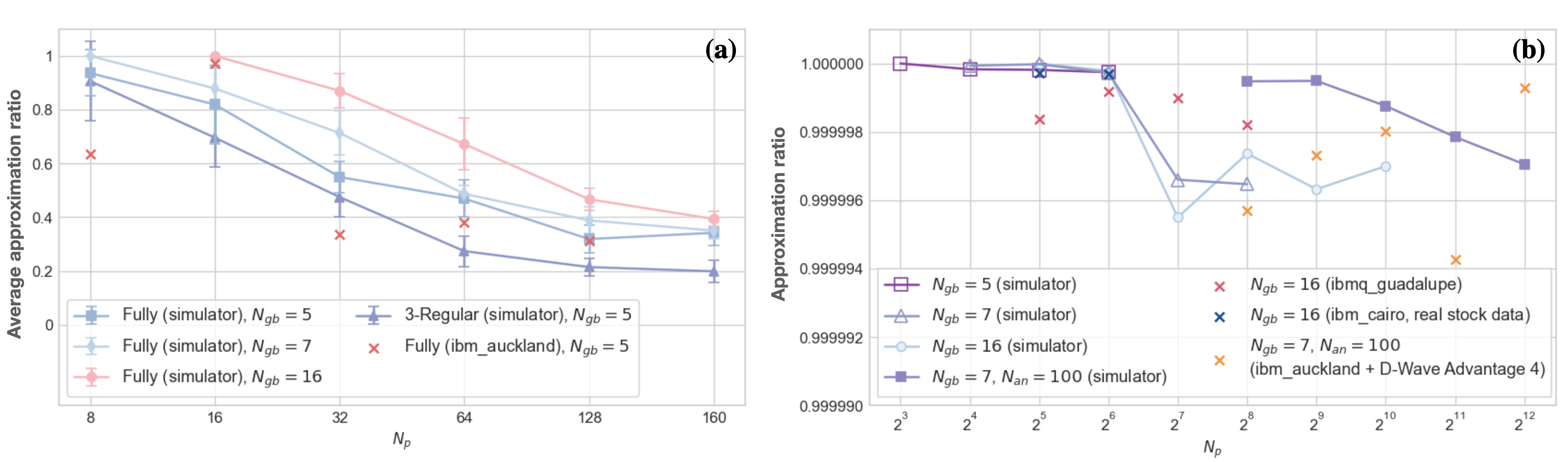}
\caption{
(a) Average approximation ratios for the fully connected random Ising problems (labeled as ``Fully") with $N_\textrm{gb} = \{5, 7, 16\}$ and for the $3$-regular random Ising problems (labeled as ``3-Regular'') with $N_\textrm{gb} = 5$ and $N_p$ from 8 to 160. We test $100$ random problems for each data point by quantum simulators with $N_s = \lceil N_p/N_{\text{gb}} \rceil$ for each random Ising problem.
For real hardware, a random Ising problem is tested for each $N_p$ with  $N_\textrm{gb} = 5$ and $N_s = \lceil N_p/5 \rceil$. (b) Approximation ratio for the portfolio optimization problems with $N_p$ from 8 to 4096. The value of $N_\textrm{gb} = \{5, 7, 16\}$  indicates that $N_s = \lceil 2N_p / N_\textrm{gb} \rceil \le N_s^\text{max} = 2^{N_\textrm{gb}}$, and thus it is possible to use a single $N_\textrm{gb}$-qubit gate-based quantum chip for the problem with result labeled with the single value of $N_\textrm{gb}$. The label with $N_\textrm{gb} = 7$ and $N_\textrm{an} = 100$ means a hybrid structure such that the subsystem problems can be solved by an $N_\textrm{an}$-qubit annealing chip and the amplitude optimization can be done by an $N_\textrm{gb}$-qubit gate-based quantum computer, with $N_s = \lceil 2N_p / N_\textrm{an} \rceil \le N_s^\text{max} = 2^{N_\textrm{gb}}$.
}
\label{fig:simu_ar_3}
\end{figure*}

\begin{figure*}
\includegraphics[scale=0.32]{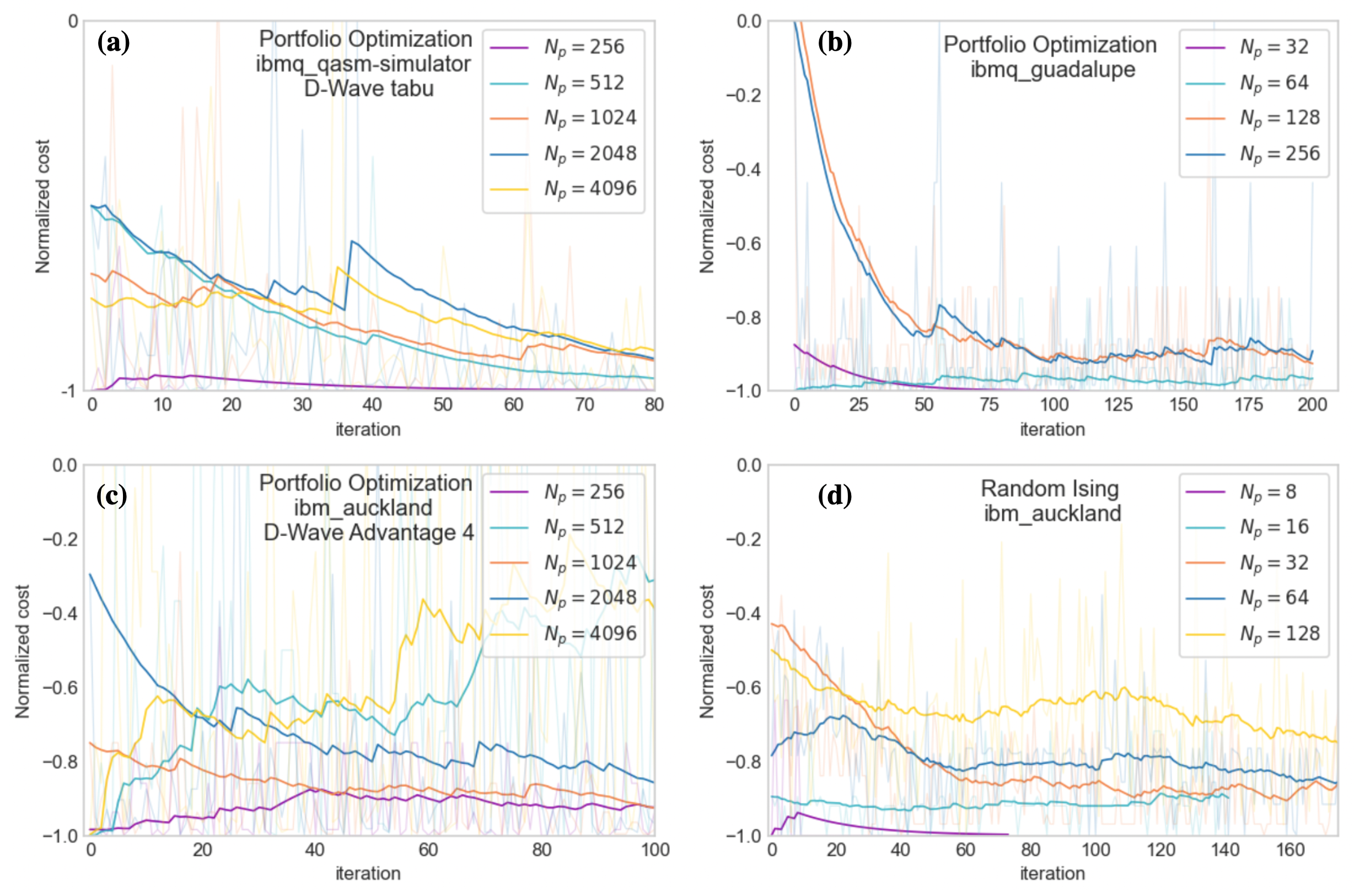}
\caption{
 Normalized cost during the optimization process for different cases: (a) Portfolio optimization with a hybrid computation structure calculated by simulators, corresponding to the solid purple square in Fig.~\ref{fig:simu_ar_3}(b). (b)Portfolio optimization with a gate-based computation structure calculated by \textsf{ibmq\_guadalupe}, corresponding to the red crosses in Fig.~\ref{fig:simu_ar_3}(b). (c)Portfolio optimization with a hybrid computation structure calculated by  \textsf{ibm\_auckland} and \textsf{D-Wave Advantage 4}, corresponding to the orange crosses in Fig.~\ref{fig:simu_ar_3}(b). (d) Random Ising problems gate-based computation structure calculated by \textsf{ibm\_auckland}, corresponding to the red crosses in Fig.~\ref{fig:simu_ar_3}(a). Data are smoothed in order to better observe the behavior, while the original data are faded in the background.
}
\label{fig:simu_ar_3_1}
\end{figure*}

In practical terms, the number of qubits of a quantum machine or computer is fixed, and thus the subsystem size to be solved by a quantum machine should also be fixed. It is crucial to find out how the performance of LSSA could be affected when solving different sizes of problems with fixed $N_g$. Considering the case that we only have access to a $N_\textrm{gb}$-qubit gate-based quantum computer, which can be used to solve both the subsystem problems with $N_g = N_\textrm{gb}$ and the amplitude optimization problems of the subsystems up to $N_s^\text{max} = 2^{N_\textrm{gb}}$. Again, this thought is tested by solving the subsystem problems with \textsf{Dwave-tabu} and then by using the \textsf{ibmq\_qasm-simulator} to do the amplitude optimization part. 
Besides the fully connected random Ising problems, 3-regular random Ising problems are also tested. A 3-regular random Ising problem is the case that each spin at a site interacts with other 3 spins at random sites with random coupling strengths, and thus the number of couplings are significantly smaller than the fully connected case.
The ranges of the random couplings $J_{ij}$ and biases $h_{i}$ of the $3$-regular random Ising problems are also both constrained in the range of $(-1,1)$.  
The approximation ratio here is defined as $R_\textrm{ar} \equiv \frac{\textrm{LSSA GSE}}{\textsf{Dwave-tabu }  \textrm{GSE}}$.
The average approximation ratios of the fully connected random Ising problems (labeled as ``Fully") with $N_\textrm{gb} \in \{5,7,16\}$ and the $3$-regular random Ising problems (labeled as ``3-Regular") with $N_\textrm{gb} = 5$ are tested for $N_p$ from 8 to 160 (note that $160 = 5 \times 2^5$). The corresponding result is shown in Fig.~\ref{fig:simu_ar_3}(a), where $100$ random problems are tested for each data point with $N_s = \lceil N_p/N_\textrm{gb} \rceil$. 
Interestingly, the results of the 3-regular random Ising problems are worse than those of the fully connected cases. This may result from the subsystem sampling process that could select configurations with interactions (couplings) that do not exist in the chosen random Hamiltonian (since we only have 3 of those couplings for each spin site), and thus the LSSA approximation in this case acts differently from that of the fully connected case.  
As shown in Fig.~\ref{fig:simu_ar_3}(a), for a fixed $N_p$, the average approximation ratios
 with larger $N_\textrm{gb}$ performing better is consistent with the results in Fig.~\ref{fig:simu_ar_2}, and 
the results are above 0.8 for $2N_\textrm{gb} \ge N_p $ in all of the fully connected cases run on simulators. 
However, for a given $N_\textrm{gb}$, the average approximation ratio decreases with increasing $N_p$ and becomes smaller than 0.5 for large $N_p$. In future work, we will develop a subsystem constructing method that depends on the coupling strengths between different sites, and this could improve the decreasing rate with $N_p$ for the average approximation ratio over the subsystem sampling scheme presented here.
We will investigate in the next subsection a specific class of the QUBO problems for which 
LSSA could have rather good performance.



\subsection{Portfolio Optimization Problems}

The random Ising problem is considered to be a general case of the problems we may encounter when dealing with quantum optimization problems. To investigate the applicability of LSSA to specific problems, we use portfolio optimization problems as our examples. A portfolio optimization problem can be also formulated into a QUBO problem \cite{portfolio2, portfolio3}, which can be further mapped to an Ising problem. We consider the portfolio optimization problem in the form \cite{portfolio1}
\begin{equation}
H = -\mu^T \omega + \frac{\gamma}{2} \omega^T \Sigma \omega + \rho (1^T \omega -K)^2 ,
\end{equation}    
where $\omega$ is an $N_p$-dimensional vector of binary decision variables, $\mu$ is the expected return vector and $\Sigma$ is the convariance matrix of size $N_p \times N_p$. The term $\mu^T \omega$ represents the return and $\frac{\gamma}{2} \omega^T \Sigma \omega$ denotes the risk term with non-negative parameter $\gamma$. The penalty $\rho$ is followed by a description of the constraint with total investment $K$. Here we can define $\text{Volatility} \equiv \sqrt{ \omega^T \Sigma \omega}$ and $\text{Sharpe ratio} \equiv \frac{\mu^T \omega}{\text{Volatility}}$. In this section, we use the simulated stock data to test LSSA on the portfolio problem, which consists of $N_p$ assets over 30 days, with parameters $\gamma = 1$, $\rho = 10N_p$ and $K = N_p/2$. For each data point solved by simulators in Fig.~\ref{fig:simu_ar_3}(b), the best result of its 3 attempts is picked while the problem instance is fixed. Unlike the case of the random Ising problems for which we generate a lot of random instances and average their results, 
in the portfolio optimization, we deal with the same dataset and only consider different numbers of assets, which may make more sense if we consider to apply our method to the real stock market (see Sec.~\ref{subsubsec:usstock}). 

Firstly, by assuming only a $5$-qubit gate-based quantum simulator ($\textsf{ibmq\_qasm-simulator}$) available, the portfolio optimization problems with $N_p \in \{8,16,32,64\}$ are tested using a single-layer QAOA with 5 iterations of the Nelder-Mead optimization to solve the subsystem problems, and the amplitude optimization process is the same as that in the random Ising case, where $N_s = \lceil 2N_p/5 \rceil$. The number of measurement shots for both the subsystem problem solving and the amplitude optimization are $8192$. The corresponding result can be found in Fig.~\ref{fig:simu_ar_3}(b). 
Interestingly, unlike the unusable small average approximation ratio obtained in Fig.~\ref{fig:simu_ar_3}(a) for the random Ising problems, LSSA holds up the approximation ratio close to 1 pretty well for the portfolio optimization problems with $N_\textrm{gb} = 5$ shown in Fig.~\ref{fig:simu_ar_3}(b), and here the approximation ratio is defined as $R_\textrm{ar} \equiv \frac{\textrm{LSSA GSE}}{\textsf{Dwave-tabu }  \textrm{GSE}}$, measuring the performance with respect to the classical  \textsf{Dwave-tabu} solver.
Result shows that the obtained cost function value is quite similar to that obtained directly by \textsf{Dwave-tabu}, with the approximation ratio $\approx 1$, almost negligibly decreasing when scaling $N_p$ up to 64. 
The effect of different numbers of measurement shots are also investigated in Appendix \ref{sec:eodvqe}. We find the results of the approximation ratio shown in Fig.~\ref{fig:simu_ar_4}(d) for the portfolio optimization problems are not that sensitive to the numbers of measurement shots, and  similar behavior for the random Ising problems can be observed in Fig.~\ref{fig:simu_ar_4}(c).

We also test $N_\textrm{gb} = 7$ for $N_p \in \{16, 32, 64, 128, 256\}$ and $N_\textrm{gb} = 16$ for $N_p \in \{32, 64, 128, 256, 512, 1024\}$ with $N_s = \lceil 2 N_p / N_{\text{gb}} \rceil$ for the both cases, and the results are also shown in Fig.~\ref{fig:simu_ar_3}(b). Similar behavior of the approximation ratio $\approx 1$ is observed while a slightly larger decrease occurs for larger values of $N_p$. Note that $N_\textrm{gb} = \{5, 7, 16\}$ are selected based on the qubit numbers of the existing quantum computers provided by IBM Quantum. Next, as mentioned in Sec.~\ref{sec:ao}, it is possible to solve the subsystem problems with $N_\textrm{an}$-qubit annealing chip and optimizing the amplitudes of the solutions of the subsystem problems by $N_\textrm{gb}$-qubit gated-based quantum computer. We simulate this case with $N_\textrm{gb} = 7$ and $N_\textrm{an} = 100$ by solving the portfolio problems for $N_p \in \{256, 512, 1024, 2048, 4096\}$ and $N_s = \lceil 2N_p/N_\textrm{an} \rceil$. For this case, the normalized cost, which is the cost function divided by the absolute value of the minimum cost, is shown in Fig.~\ref{fig:simu_ar_3_1}(a). We notice that the cost function saturates to a small value during the optimization process quite quickly, indicating that the initial state of the solved subsystem problems with random initialized coefficients constitutes a good initial guess. \revise{}

\section{Calculations by Gate-Based Quantum Computers and Quantum Annealers}
\label{sec:gbcqa}
\subsection{Random Ising Problems}
After the investigation with the calculations performed by simulators, 
we use the gate-based quantum computers provided by IBM and the quantum annealer from D-Wave to perform the calculations to show the capability of the LSSA algorithm on the currently available quantum computing machines.
For the subsystem problems solved by the gate-based quantum computers, a single layer QAOA is used with 5 iterations of the Nelder-Mead optimization, and 8192 measurement shots are used for each calculation of the expectation value. 
The results of the random Ising problem tested by \textsf{ibm\_auckland} with $N_g=N_\textrm{gb} = 5$ for $N_p \in \{8, 16, 32, 64, 128\}$  and $N_s = \lceil 2N_p/5 \rceil$ are shown in red crosses in Fig.~\ref{fig:simu_ar_3}(a), and the normalized cost in this case is shown in Fig.~\ref{fig:simu_ar_3_1}(d). 
In this section, the approximation ratio is defined as $R_\textrm{ar}\equiv\frac{\textrm{LSSA GSE}}{\textsf{Dwave-tabu}  \textrm{GSE}}$ again. 
Although we only test one random Ising problem for each $N_p$, the trend of the average approximation ratio is similar to that obtained by the simulators, i.e., it decreases considerably to a low value when $N_p \gg N_g$, indicating a relatively poor performance.

\subsection{Portfolio Optimization Problems}
\subsubsection{Simulated Stock Data}

First, the gate-based quantum computer of \textsf{ibmq\_guadalupe} with  $N_g=N_\textrm{gb} = 16$ is used to solve the portfolio optimization problems for $N_p \in \{32, 64, 128, 256\}$  and $N_s = \lceil 2N_p/16 \rceil$. The results are shown in red crosses in Fig.~\ref{fig:simu_ar_3}(b). 
Similar to the results by the simulator in Fig.~\ref{fig:simu_ar_3_1}(a), the normalized cost decreases to the lowest value in very few iterations as shown in Fig.~\ref{fig:simu_ar_3_1}(b).
Then the hybrid annealing and gate-based computation structure is demonstrated by \textsf{D-Wave Advantage 4} with  $N_\textrm{an} = 100$ and \textsf{ibm\_auckland}  with $N_\textrm{gb} = 7$ for solving the portfolio optimization problems for $N_p \in \{256, 512, 1024, 2048, 4096\}$  and $N_s = \lceil 2N_p/100 \rceil$. The results are shown in orange crosses in Fig.~\ref{fig:simu_ar_3}(b). 
Note that even though \textsf{D-Wave Advantage 4} is a 5760-qubit system, the upper limit of the number of variables of an arbitrary (a fully-connected) Ising problem that can be solved by it is around 145 due to the connectivity geometry of the qubit hardware \cite{dwave_upper_limit}. 
So we use this quantum annealing machine to conservatively solve the subsystem problems with size $N_\textrm{an} = 100$, and for each subsystem problem 5000 shots are used.
The approximation ratio in this case as shown in Fig.~\ref{fig:simu_ar_3}(b) is also almost equal to 1, 
 which is quite impressive considering we only use $7 + 100$ qubits to solve the problem with size almost 40 times larger than the qubit number of our hardware. However, due to our definition of the approximation ratio used here, the good performance just means the result is very close to that obtained from the \textsf{Dwave-tabu} solver,  and is not guaranteed to be close to the actual solution of the problem.

\subsubsection{US Stock Market Data}
\label{subsubsec:usstock}

\begin{figure*}
\includegraphics[scale=0.3]{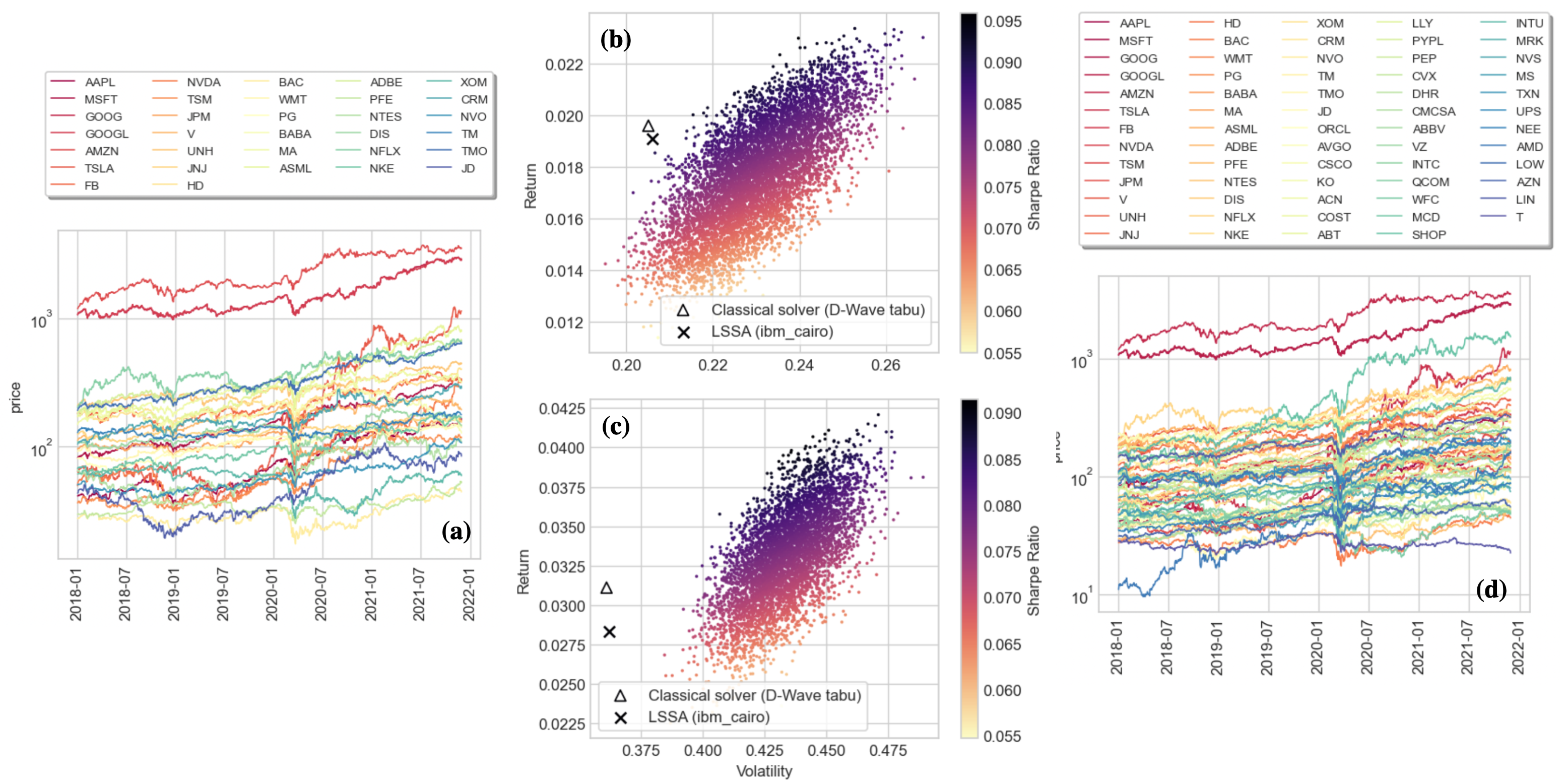}
\caption{Data and results of the Sharpe ratio of the the portfolio optimization problem
for the US stock market. (a) US stock market data for the stocks of $N_p = 32$ largest companies sorted by market cap. (b) Return and volatility of the Sharpe ratios of the portfolio optimization problem with $N_p = 32$ stocks, obtained from the classical solver (triangle) and LSSA (cross). The scattered dots represent the results from random sampling of $5 \times 10^3$ combinations of the portfolios.
(c) Similar to (b) but for $N_p = 64$ stocks. (d) Similar to (a) but for $N_p = 64$ stocks.
} 

\label{fig:portfolio_us_stock}
\end{figure*}

\begin{figure}
\includegraphics[scale=0.16]{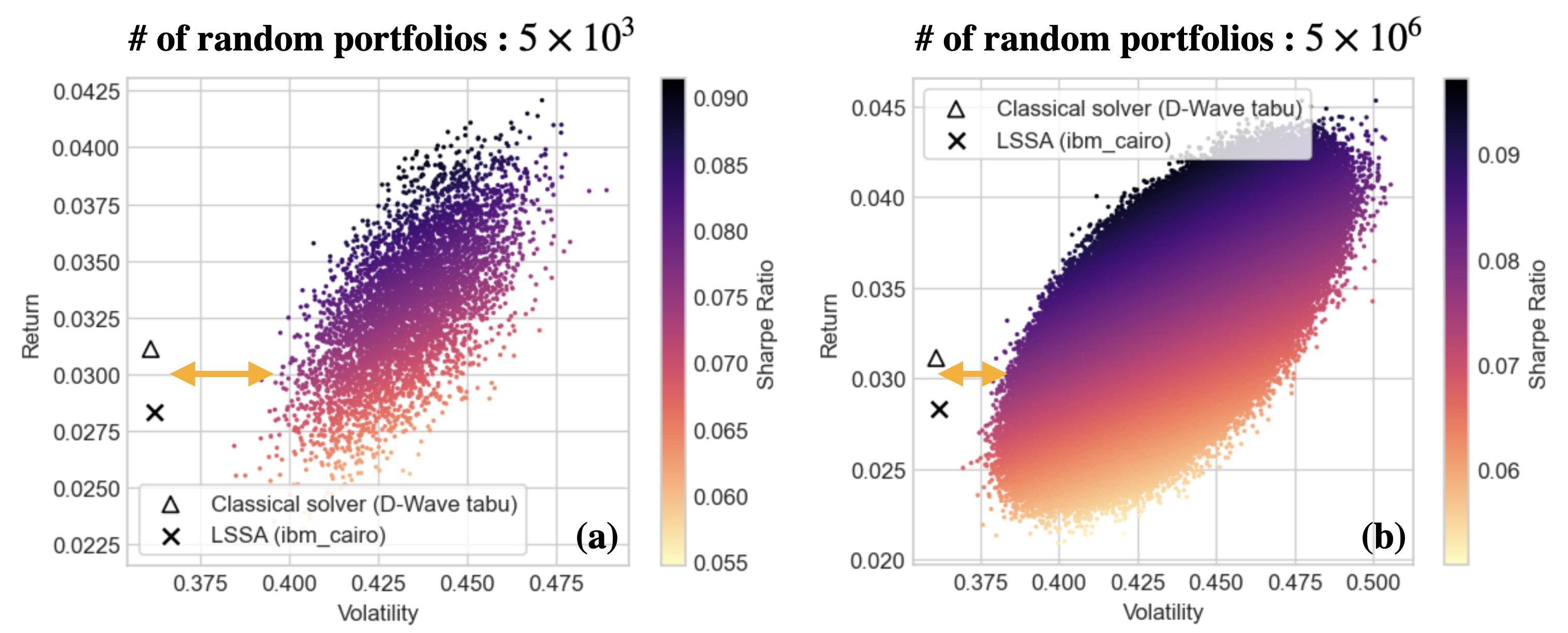}
\caption{Return and volatility for different numbers of random portfolio compositions 
(a) $5 \times 10^3$ and (b) $5 \times 10^6$.
The distance between the results obtained from the classical solver (triangle) and LSSA (cross) and the results from the random sampling portfolio compositions is indicated by an orange double-arrow line.
} 

\label{fig:portfolio_us_stock_sample}
\end{figure}

To examine the capability of LSSA for portfolio optimization in real world, we also test the real data from the US stock market \cite{yahoostock} with a gate-based machine for subsystem problem size $N_\textrm{gb} = 16$ to solve the portfolio optimization problems of $N_p \in \{32, 64\}$,  and  $N_s = \lceil 2N_p/16 \rceil$. For each problem, we pick stocks of 
$N_p$ largest Companies by market cap
and use data over 47 months to construct our problem Hamiltonian, where parameters $\gamma$, $\rho$ and $K$ are the same as the above cases. The US stock data for $N_p = 32$ and $N_p = 64$ are shown 
in Figs.~\ref{fig:portfolio_us_stock}(a) and \ref{fig:portfolio_us_stock}(d), respectively. 
Both the subsystem solving part and the solution amplitude optimization are done by the IBM Quantum machine \textsf{ibm\_cairo}. 
The result of the approximation ratio is shown in blue crosses in Figs.~\ref{fig:simu_ar_3}(b).  
The Sharpe ratios obtained by both the classical solver (\textsf{Dwave-tabu}) and LSSA with  \textsf{ibm\_cairo} for $N_p = 32$ and $N_p = 64$ shown in Fig.~\ref{fig:portfolio_us_stock}(b) and \ref{fig:portfolio_us_stock}(c), respectively,
are significantly larger than those from random stock portfolio compositions [scattered dots in Figs.~\ref{fig:portfolio_us_stock}(b) and \ref{fig:portfolio_us_stock}(c)], but the Sharpe ratios from LSSA are slightly lower than those from the classical solver in both $N_p = 32$ and $N_p = 64$ cases. 
One may wonder 
why the points of both the optimized solutions from the classical solver and LSSA do not fall within but instead are far from the random sampling points (scattered dots) of the stock portfolio compositions.
This is because the number of possible portfolio compositions that satisfy our constraint is $C(64, 32) = \frac{64!}{32!32!}\approx 1.8 \times 10^{18}$, and the number of random sampling portfolio compositions of $5 \times 10^3$ is only a very tiny portion of the full solution space.
We show in Fig.~\ref{fig:portfolio_us_stock_sample} that with a larger number of random portfolio compositions of $5 \times 10^6$, the distribution of the scattered points becomes wider. As a result, our optimized solutions becomes closer to the distribution of the results from the random portfolio compositions. Even though the number of random sampling points increases by a factor of $10^3$, it is still much smaller than the number of total possible portfolio compositions in the solution space. This also highlights the difficulty of trying to find the optimal solution of the combinatorial optimization problem
through the random sampling search or the brute-force or exhaustive search algorithm.   
It is thus reasonable that the optimized solutions are distant from the distribution of the results obtained from the random portfolio compositions.


\section{Discussions}
\label{sec:discuss}

Although the idea to reduce the required number of qubits to solve a larger-size problems seems compelling, the fact that the subsystem sampling method of LSSA neglects some of the couplings between variables (spin sites) makes it not applicable to all kinds of Ising problems. For the random Ising problems, the average approximation ratio (see Fig.~\ref{fig:simu_ar_1} and Fig.~\ref{fig:simu_ar_2}) could be increased by tuning up $N_g$ and $N_s$, while its scaling with $N_p$ is in a declining trend. In this case, LSSA could be useful when $(N_p / N_g) \sim O(10^0)$, i.e., when the number of qubits of the quantum machine is not far from the size of the target problem. If $(N_p / N_g) \sim O(10^1)$,  we find in Fig.~\ref{fig:simu_ar_3}(a) that the problem size $N_p$ is too large to have a reasonably good solution for the random Ising problems by LSSA using both the quantum simulator and real hardware (\textsf{ibm\_auckland}). However, when it comes to the portfolio optimization problems, even when $N_p / N_g \sim O(10^1)$, the average approximation ratio can be very close to 1, maintaining in the range of $1 - O(10^{-6})$ using both the simulator and real hardware (\textsf{ibmq\_guadalupe}, \textsf{ibm\_auckland}, \textsf{ibm\_cairo} and \textsf{D-Wave Advantage 4}). 
This could result from the coupling strengths are quite different for the random Ising problems and portfolio optimization problems.
The strengths of coupling $J_{ij}$ and biases $h_i$ of the random Ising problem are distributed randomly between the values $-1$ and $1$, so their strengths are comparable.
While for the portfolio optimization problem, the coupling strengths are the covariance values between different assents (stocks) in our formulation, which are mostly in the order of $10^{-2}$ much smaller than the expected returns of the assets (stocks). Thus the neglected couplings in the subsystem sampling process make much more considerable effect in obtaining the full-system ground state energy for the random Ising problem than for the portfolio optimization problem. Furthermore, the VQE subsystem amplitude optimization procedure with the full-system Hamiltonian in the final stage also tries to adjust and recover some correlation lost in the subsystem sampling. This also helps LSSA have a very high-quality approximation ratio for the portfolio optimization problems as the correlation lost due to the subsystem sampling process is not that substantial.


The maximum number of $N_s$ is determined by the size of the gate-based quantum computer ${N_\text{gb}}$ for amplitude optimization of the solution of the subsystem problems, i.e., $N_s^\text{max} = 2^{N_\text{gb}}$. Therefore, the problem size $N_p$ could be scaled up to 
$N_p^\text{max}=N_\text{gb}N_s^\text{max}=N_\text{gb}2^{N_\text{gb}}$ for a gate-based machine of size ${N_\text{gb}}$, i.e., each spin site should be at least sampled once.  
Since there is no causality between solving different subsystem problems in LSSA, it is possible to solve the subsystem problems in parallel if we have multiple quantum computers of the same or larger sizes.  
On the other hand, one can in a hybrid structure take the advantage of a larger effective number of qubits of quantum annealing machines to solve subsystem Ising problems of a larger size (i.e., $N_\text{an} > N_\text{gb}$), and then can use a gate-based quantum computer to encode and optimize the subsystem solution coefficients in the Hilbert space to find the solution of a full-system problem of a larger size up to $N_p^\text{max}=N_{\text{an}}2^{N_{\text{gb}}}$.
Alternatively, one can also take the advantage of $N_\text{an} > N_\text{gb}$
to decrease the number of $N_s$ to $\tilde{N_s} = \frac{N_\text{gb}}{N_\text{an}} N_s$ when keeping the full-system-problem size fixed. 
For the result demonstrated in Fig.~\ref{fig:simu_ar_3}, 
we do not solve the full-system problem of the maximum size $N_p^\text{max}$. 
Instead, we only solve problems of size $N_p<N_p^\text{max}$ with the number of the subsystems $N_s<N_s^\text{max}$ in order
to have reasonable execution times to demonstrate the behavior of LSSA by our current hardware. This could be conquered by solving the subsystem problems in parallel as mentioned above if we would have many small-size quantum machines. 

We next discuss the computational properties of LSSA. 
For the case that the decrease of the performance of the approximation ratio is negligible (e.g. for the portfolio optimization problems), by fixing the problem size $N_p$, when the subsystem size $N_g$ decreases, a larger number of the randomly sampled  subsystem problems $N_s$ is apparently required, that is, $N_s \propto N_g^{-1}$. Compared to directly solving the problem by a $N_p$-qubit quantum computer, LSSA needs an additional time to solve the subsystem problems before executing the VQE algorithm for amplitude optimization  that involves the calculation of the full-Hamiltonian cost function. Thus we can describe the time requirement of LSSA as $t_{\text{LSSA}} = N_s t_{\text{sub}} + t_{\text{VQE}}$, where $t_{\text{sub}}$ is the time to solve for a subsystem problem and $t_{\text{VQE}}$ is the time to run the VQE algorithm for amplitude optimization. Since $t_{\text{VQE}}$ requires a polynomial time (considering only the Hamiltonian with polynomially growing sum of independent observables \cite{vqe1, polynovqe1}) and $N_s \propto N_p$, LSSA is still an polynomial time algorithm with less qubit requirement. 

So far, we have shown that a larger-size problem can be approximately solved by dividing it into problems of small-size subsystems.  We call this LSSA a level-1 approximation as only one level of subsystem reduction is made. However, this procedure is not constrained to be only done once. To further extends the concept of LSSA, we can even divide the subsystems into further smaller-size sub-subsystems, which may be useful when multiple quantum computers with a small system size are available, such that all sub-subsystem problems can be calculated in parallel. We called this case a level-2 approximation. In Appendix \ref{sec:L2}, we show how a 5-qubit gate-based quantum computer is possible to approximately solve  for the portfolio optimization problem with up to 5120 assets (stocks), yet still yielding pretty good performance.

\section{Conclusion and Outlook}
\label{sec:outlook}

Targeting to tackle the Ising problems of a size much larger than that of our current hardware, we propose a new algorithm, LSSA, to sample the full-system problem into a number of smaller-size subsystems. The subsystem problems are then solved by the subsystem solvers of gate-based or annealing quantum machines. After that, the contributions of the solutions of different subsystems are considered as amplitudes in a VQE optimization algorithm using the full-system Hamiltonian, and the optimal contribution of each subsystem is then reconstructed to determine the approximate full-problem ground state configuration.
We solve the random Ising problems with sizes up to 160 variables using 5 qubits in a gate-based quantum computer, and solve the portfolio optimization problems with sizes up to 4096 variables using 100 qubits in a quantum annealer and 7 qubits in a gate-based quantum computer using the level-1 approximation of LSSA.
Our result gives an example to scale up the capability of the current quantum computing resources by combining different computational architectures of quantum machines. 
We also show the extendibility of LSSA to a deeper level of approximation by employing the level-2 approximation of LSSA for the portfolio optimization problems up to 5120 ($N_{\text{gb}}2^{2 N_{\text{gb}}}$) variables with pretty good approximation ratio values using only 5 ($N_{\text{gb}}$) qubits in a gate-based quantum computing scenario.
The portfolio optimization problems could be excellent problems for LSSA as the exact ground-state solutions are desired but not required as long as for each problem, high-quality solutions close to the ground-state solution can be found. After all, one may not expect to have a portfolio composition that can give the maximum return and the minimum risk for every investment, but will be happy as long as one can make a substantial profit. 
Although the quality of the performance of LSSA could be problem-dependent, 
the future work aims to find what kinds of problems are suitable for LSSA and to improve the subsystem sampling effectiveness and efficiency by considering a coupling-dependent sampling method.


\acknowledgments{We thank 
IBM Quantum Hub at NTU for providing computational
resources and accesses for conducting the gate-based quantum
computer experiments,
 and the Center for Quantum Frontiers of Research and Technology, National Cheng Kung University, Taiwan for providing accesses to Amazon Braket. 
We also acknowledge the support from MediaTek Inc., Taiwan.
H.S.G. acknowledges support from the the Ministry of Science and Technology, Taiwan under Grants No.~MOST 109-2112-M-002-023-MY3, 
No.~MOST 109-2627-M-002-003, No.~MOST 107-2627-E-002-001-MY3, No.~MOST 111-2119-M-002-006-MY3,
No.~MOST 109-2627-M-002-002, No.~MOST 110-2627-M-002-003,
No.~MOST 110-2627-M-002-002, No.~MOST 111-2119-M-002-007, No.~MOST 109-2622-8-002-003, and 
No.~MOST 110-2622-8-002-014,
from the US Air Force Office of Scientific Research under
Award Number FA2386-20-1-4033,
and from the
National Taiwan University under Grants
No.~NTU-CC-110L890102 and No.~NTU-CC-111L894604, 
H.S.G. is grateful to the support 
from the Physics Division, National Center for Theoretical Sciences, Taiwan.}

\appendix

\begin{figure*}
\includegraphics[scale=0.33]{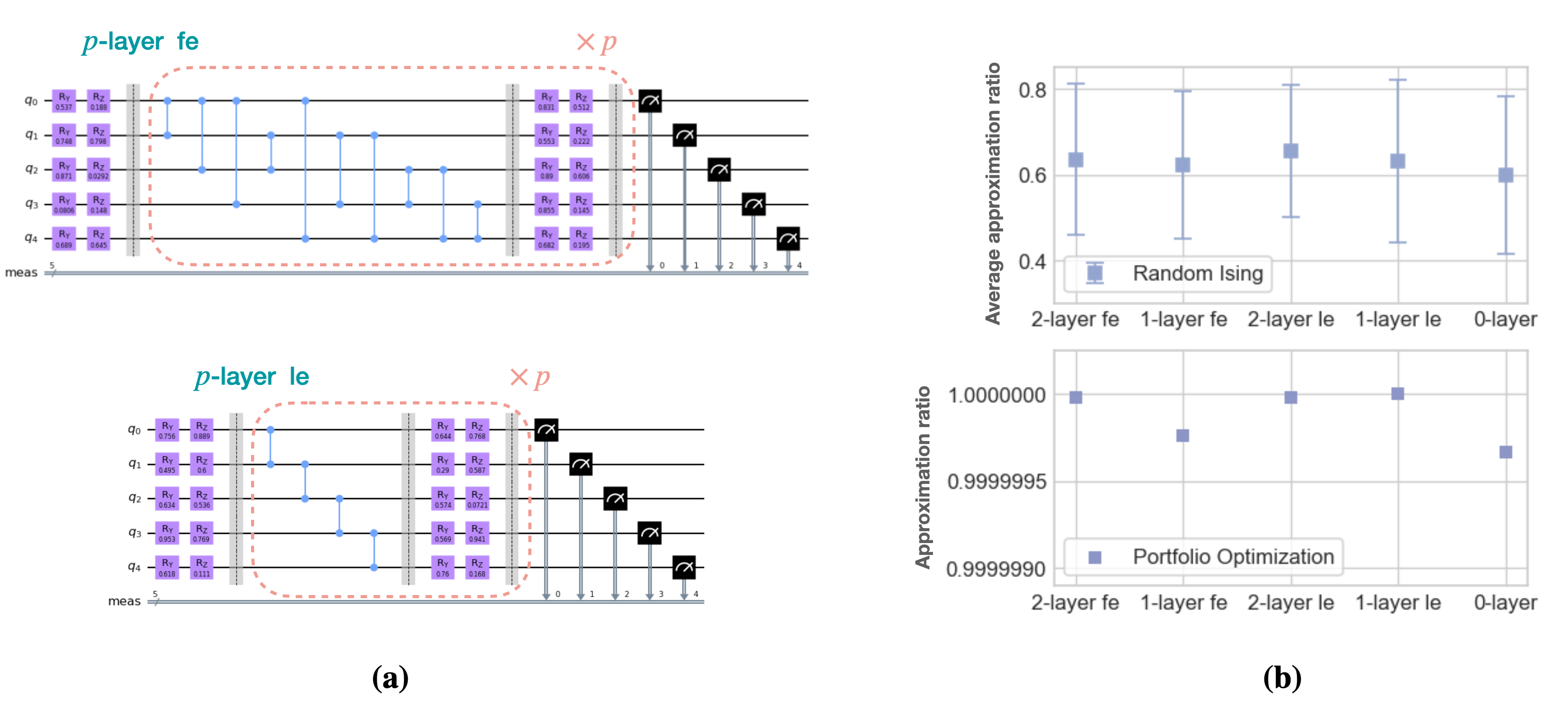}
\caption{Effects of different VQE circuit structures (ansatzes) for the subsystem amplitude optimization. (a) Visualization of the VQE circuits. The $p$-layer structure means that the circuit in the dashed box is repeated $p$ times, and labels ``fe'' and ``le'' indicate the entanglement layer being full entanglement (the upper graph) and linear entanglement (the lower graph), respectively. Note that 0-layer means the circuit without the dashed box. (b) Average approximation ratio of the random Ising problems and approximation ratio of the portfolio optimization problems with $N_p = 20$, $N_g = 5$ and $N_s = 8$. For each data point of the random Ising problems, 100 problems are solved and averaged. In the portfolio optimization plot, the best result of their 10 attempts for the same problem is picked. }
\label{fig:eodvqe}
\end{figure*}

\begin{figure*}
\includegraphics[scale=0.28]{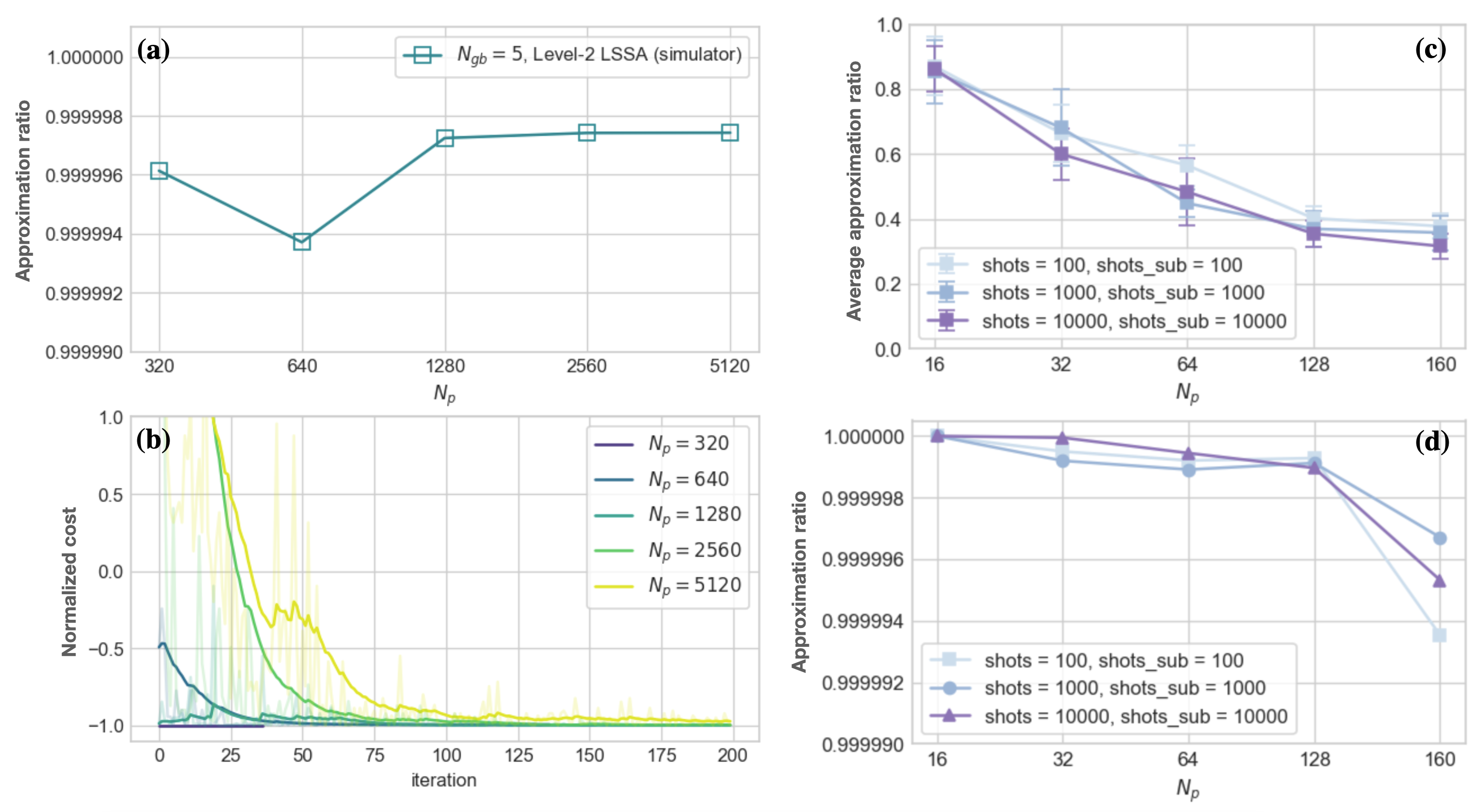}
\caption{(a) Approximation ratio of the portfolio optimization problems up to $N_p = 5120$ solved by \textsf{ibmq\_qasm-simulator} using 5 qubits. For each data point, the best result among its 3 attempts is picked. (b) Normalized cost during the optimization process for different values of $N_p$ from $N_p = 320$ to $N_p = 5120$. The data are smoothed to better observe the behavior, while the original data are faded in the background. (c) Average approximation ratios of the random Ising problems for different measurement shots by \textsf{ibmq\_qasm-simulator}, with $N_g = 5$, $N_s = \lceil N_p/5 \rceil$ and $N_p \in \{16,32,64,128,160 \}$. 100 problems are tested for each data point. (d) Approximation ratios of the portfolio optimization problem for different measurement shots by \textsf{ibmq\_qasm-simulator}. Parameters $N_g$, $N_p$ and $N_s$ are the same as those in (c), but for each data point, the best result among its 20 attempts is picked.}
\label{fig:simu_ar_4}
\end{figure*}

\section{Effect of different circuit structures of VQE}
\label{sec:eodvqe}

The VQE circuit structure (ansatz) used in the subsystem amplitude optimization stage is also one of the hyperparameters that may affect our result. 
The structure of the quantum circuit of our VQE 
consists of alternating entanglement layers and rotation layers as shown in Fig.~\ref{fig:vqe} and Fig.~\ref{fig:eodvqe}(a).
The $p$-layer structure in Fig.~\ref{fig:eodvqe} means that the circuit in the dashed box is repeated $p$ times, and labels ``fe'' and ``le'' indicate that the entanglement strategies of the entanglement layers in the dashed box being full entanglement [the upper graph in Fig.~\ref{fig:eodvqe}(a)] and linear entanglement [the lower graph in Fig.~\ref{fig:eodvqe}(a)], respectively.
In the paper, our main result is computed by the ``2-layer fe'' structure described in Fig.~\ref{fig:vqe} and Fig.~\ref{fig:eodvqe}(a). 
In this appendix, we tune the type of the entanglement strategies and the number of the parameterized circuit layers in the dashed box for solving the problems with $N_p = 20$, $N_g = 5$ and $N_s = 8$ by exact diagonalization for subsystems and \textsf{ibmq\_qasm-simulator} for amplitude optimization. For the random Ising problem, 100 problems are solved and averaged, while for the portfolio optimization problem, we pick the best result of their 10 attempts for the same problem. As shown in Fig.~\ref{fig:eodvqe}(b), the average approximation ratios for the cases with linear entanglement are slightly better than those with full entanglement in simulations on simulators. Here the approximation ratio is defined as $R_\text{ar}=\frac{\text{LSSA GSE}}{\text{Exact GSE}}$. 
For the problem we consider, increasing the number of parameterized circuit layers help slightly the performance in the average approximation ratios for both the cases with full entanglement and linear entanglement, but the effect is less obvious for the cases with linear entanglement. Both the quantum circuits with linear entanglement and full entanglement layers have comparable performance. But if only one rotation layer and no entanglement layer 
is used for the circuit, then the performance in the average approximation ratios degrades.

\section{Level-2 approximation}
\label{sec:L2}

Dividing the full problem into subsystems makes it possible to solve for the problem with size larger than that of our quantum hardware. Moreover, it may be possible to decompose the subsystems into sub-subsystems, and this will significantly increase the maximal problem size we can solve with our available quantum hardware.
Let us call the procedure that use LSSA once the level-1 approximation 
since the problem is decomposed once and the subsystems are in the same size.  In the level-1 approximation, an $N_\text{gb}$-qubit gate-based can solve QUBO problems up to $N_\text{gb} 2^{N_\text{gb}}$ variables. If we further consider $N_\text{gb} 2^{N_\text{gb}}$ as our subsystem size, with $2^{N_\text{gb}}$ subsystems that can be encoded into VQE for amplitude optimization, we can solve QUBO problems with sizes $N_p = N_\text{gb} 2^{2N_\text{gb}}$ in total. We call this further decomposition the level-2 approximation.
This means that with only 5-qubit gate-based quantum computers, the maximal size of the problems that can be solve in the level-2 approximation is $N_p = 5 \times 2^{10} = 5120$.   Again, this means that we will have $2^{2N_\text{gb}}$ subsystems to be solved before dealing with the full problem Hamiltonian for the VQE amplitude optimization. This could be a quite large amount and could make LSSA too slow to be useful. However, this may be conquered by solving the subsystem problems on several quantum computers in parallel, due to the fact that there is no causality between the subsystems in the same level. In practical terms, it would be much easier to have hundreds or dozens of very reliable 5-qubit or 10-qubit quantum computers than 
to have or develop a single reliable quantum computer with several thousands qubits. 
Quantum computer clusters utilizing the concept of LSSA could be one of the possible scenarios that the real-world use case of quantum computing applications could be found in the future. 
To demonstrate what level-2 LSSA is capable of, 
in this appendix, we use it to solve, as an example, the portfolio optimization problems with stock data sizes $N_p \in \{320, 640, 1280, 2560, 5120\}$. 
With 5-qubit quantum computers available, 
the subsystem sizes $N_\text{gb}^{(1)} = 160$ and $N_\text{gb}^{(2)} = 5$ in the level-1 and level-2 approximations mean that the first level of LSSA divides the problem into subsystem problems with sizes $N_\text{gb}^{(1)} = 160$, then the second level of LSSA further divides the problem from subsystem size $N_\text{gb}^{(1)} = 160$ to $N_\text{gb}^{(2)} = 5$, with $N_s^{(1)} = \lceil N_p / N_\text{gb}^{(1)} \rceil$ and $N_s^{(2)} = 2^{N_\text{gb}^{(2)}}=32$. In Fig.~\ref{fig:simu_ar_4}, we show the approximation ratio and normalized cost during the optimization process calculated by simulators for problem sizes mentioned above. The approximation ratio here is also defined as $R_\text{ar}=\frac{\text{LSSA GSE}}{\textsf{Dwave-tabu }  \text{GSE}}$. It is interesting to observe that the approximation ratio still holds up well in the portfolio optimization problems even if we use level-2 LSSA.


\nocite{*}
\bibliography{LSSA}


\end{document}